\documentstyle[12pt]{article}

\begin{document}
\thispagestyle{empty}

\begin{center}
\LARGE \tt \bf{Nongeodesic motion of spinless particles in the teleparallel gravitational wave background}
\end{center}

\vspace{1cm}

\begin{center} {\large L.C. Garcia de Andrade\footnote{Departamento de
F\'{\i}sica Te\'{o}rica - Instituto de F\'{\i}sica - UERJ

Rua S\~{a}o Fco. Xavier 524, Rio de Janeiro, RJ

Maracan\~{a}, CEP:20550-003 , Brasil.

E-Mail.: garcia@dft.if.uerj.br}}
\end{center}

\vspace{1.0cm}

\begin{abstract}
The motion of spinless particles in Riemann-Cartan (RC) $U_{4}$ and teleparallel spacetimes is revisited on the light of gravitational waves hitting on the spinless test particles. It is shown that in the case Cartan contortion is totally skew-symmetric the spinless particles follow geodesics but if this symmetry is dropped they are able to follow nongeodesic worldlines in torsionic background. The case of totally skew contortion and spinning particles may appear either in $T_{4}$ or $U_{4}$ spacetimes. We consider $T_{4}$ nongeodesic motion of spinless test particles in the field of gravitational waves (GW). It is shown that the general contortion obeys a tensor wave constraint and the analysis of a ring of spinless test particles as in General Relativity (GR) hit by the GW leads to damping contortion effects on the nongeodesic motion is undertaken. It is also shown that the contortion and gravitational waves possess a difference of phase of $\frac{\pi}{2}$ and a contortion at the surface of the Earth of $10^{-24} s^{-1}$ computed by Nitsch in the realm of teleparallelism is used to obtain a deviation of the lenght of the separation of test spinless particles compatible with the GR result. To obtain this result data from GW LISA detector is used.      
\end{abstract}
\vspace{1.0cm}       
\begin{center}
\Large{PACS numbers : 0420,0450.}
\end{center}

\newpage
\pagestyle{myheadings}
\markright{\underline{Nongeodesic motion in teleparallelism}}

\paragraph*{}
\section{Introduction}
Recently a great interest in Einstein-Weitzenb\"{o}ck \cite{1} teleparallelism and GW in Riemann-Cartan spacetime \cite{2,3} has been put forward. On the teleparallel gravity side in order to investigate the motion of spinless particles in teleparallel gravity and also to solve the old GR problem of the complex of energy density and pseudotensor of energy-momentum density by de Andrade, Guillen and Pereira \cite{4}. Earlier Hojmann \cite{5} has discuss the equation of motion of spinning particles in torsion theories by making use of a Lagrangean formalism. In this paper Hojmann investigates the motion of spinning particles by building  constant of motion which depends on the interaction between contortion tensor and the tensor of angular-momentum of spinning particles which reduces to the constant of motion of GR in the absence of torsion or spinning particles. Nevertheless contrary to general belief is possible to show that not only spinning particles can interact with torsion as showed by Hehl \cite{6}, but also interacts with spinless particles as shown by Kleinert \cite{7} by making use of a variation of the action taking into account the variation of extrema. In this paper we perform a detailed investigation of the problem of motion of spinless test particles on the contortion and GW background . Earlier Nieto and Ryan  \cite{8} investigated the motion of spinning particles in the front-wave of a GW in the context of GR ,thus in a certain sense this paper complements the Nieto-Ryan paper. This paper is organized as follows. In the section II we review the use of geodesic deviation equation derived by Hehl \cite{6} applied to the spin and spinless particles in $T_{4}$ and $U_{4}$ to show that in the case contortion is totally skew-symmetry the spinless particles follow geodesic worldlines in $T_{4}$ while  spinning  particles in $T_{4}$ follows nongeodesic motion even in $T_{4}$. However, these results are not only due to the nature of spin but due to the nature of spacetime. For example we also show that the same spinless particles when the contortion is not totally skew follows nongeodesic motion even in the case of $T_{4}$. The case of spinning particles is well-known and we do not discuss it here in much detail. For recent reviews that include the motion of spinning particles in RC $T_{4}$ spacetime and Einstein-Cartan gravity the reader is refered to the very recent reviews of Shapiro \cite{9} and Hammond \cite{10}. In section III tensor gravitational perturbation of Minkowski flat spacetime representing a plane gravitational wave passing by the ring composed of spinless particles in torsion spacetime background is investigated and we show that the presence of not totally symmetric contortion leads to a damping effect on the spinless particles motion under the action of GW. Therefore despite of the fact that GR experimental results should agree with teleparallel results up to order $O(v^{5})$ in the velocities v as shown by Smalley \cite{11} in the context of metric-affine gravity and teleparallelism one may obtain an estimete for contortion based on the well-known results of GR GW which is done in section IV. Conclusions and discussions are given in section $5$.

\section{Motion Equation for Spinless Particles in $T_{4}$}
Let us consider the equation of motion of spinning particles in $U_{4}$ as given in Nitsch \cite{12}  
\begin{equation}
\frac{Dp^{\mu}}{ds}=-\frac{1}{2}{{R^{\mu}}_{{\nu}{\alpha}{\beta}}}(\Gamma) u^{\nu} S^{{\alpha}{\beta}} -{K^{\mu}}_{{\alpha}{\beta}} u^{\alpha}p^{\beta}
\label{1}
\end{equation}
where ${{R^{\mu}}_{{\nu}{\alpha}{\beta}}}({\Gamma})$ is the Riemann-Cartan curvature tensor,${p}^{\mu}$ is the four-momentum of the spinning particle with spin angular-momentum tensor $S^{{\alpha}{\beta}}$ while $u^{\alpha}$ is the four-velocity. Here ${\alpha}=0,1,2,3$ and latin indices represents $i,j=1,2,3$. The derivative operator $\frac{D}{ds}$ represents the Riemannian absolute derivative. The relation between $u^{\alpha}$ and $p^{\mu}$ is given by 
\begin{equation}
p^{\mu}=mu^{\mu}-S^{{\mu}{\nu}}\frac{Du_{\nu}}{ds}
\label{2}
\end{equation}
where $p^{\nu}u_{\nu}=m^{2}$ and m represents the particle mass. The spin transport Mathisson equation is 
\begin{equation}
\frac{DS^{{\mu}{\nu}}}{ds}=p^{\mu}u^{\nu}-p^{\nu}u^{\mu}
\label{3}
\end{equation}
The spinless particle condition is given by $S^{{\mu}{\nu}}=0$. This hypothesis reduces the motion equations to  
\begin{equation}
p^{\mu}=mu^{\mu}
\label{4}
\end{equation}
\begin{equation}
\frac{Dp^{\mu}}{ds}= -{K^{\mu}}_{{\alpha}{\beta}} u^{\alpha}p^{\beta}
\label{5}
\end{equation}
where we already made use the $T_{4}$ teleparallel condition of the vanishing of RC curvature to obtain expression (\ref{5}). Substitution of expression (\ref{4}) into formula (\ref{5}) yields
\begin{equation}
\frac{Du^{\mu}}{ds}= -{K^{\mu}}_{{\alpha}{\beta}} u^{\alpha}u^{\beta}
\label{6}
\end{equation}
From this expression one immediatly notes that the symmetry of spacetime contortion is fundamental for the type of geodesic or nongeodesic nature of motion of the spinless particles in $T_{4}$. When the contortion is totally skew-symmetric or $K_{{\mu}{\nu}{\beta}}=K_{[{\mu}{\nu}{\beta}]}$ the RHS term of the equation (\ref{6}) vanishes and we are left with the equation of motion  of geodesics
\begin{equation}
\frac{Du^{\mu}}{ds}= 0
\label{7}
\end{equation}
while in the case contortion possess the symmetry $K_{{\mu}({\alpha}{\beta})}$ representing the 
symmetry of the last two indices allows us to observe that the spinless particles in $T_{4}$ in this case follow nongeodesics sometimes called autoparallels of RC connection. Since the absolute derivative (\ref{7}) is explicitly given by
\begin{equation}
\frac{Du^{\mu}}{ds}=\frac{du^{\mu}}{ds}+{{{\Gamma}^{0}}^{\mu}}_{{\alpha}{\beta}}u^{\alpha}u^{\beta}
\label{8}
\end{equation}
where 
\begin{equation}
{{{\Gamma}^{0}}^{\mu}}_{{\alpha}{\beta}}=\frac{1}{2}g^{{\mu}{\gamma}}(g_{{\gamma}{\alpha},{\beta}}+g_{{\gamma}{\beta},{\alpha}}-g_{{\alpha}{\beta},{\gamma}})
\label{9}
\end{equation} 
is the Riemann-Christoffel symmetric connection.
By considering the metric tensorial perturbations around the Minkowski spacetime metric ${\eta}_{{\mu}{\nu}}$ reads
\begin{equation} 
g_{{\mu}{\nu}}={\eta}_{{\mu}{\nu}}+h_{{\mu}{\nu}}
\label{10}
\end{equation}
where $|h_{{\mu}{\nu}}|<<1$ is the gravitational tensor perturbation corresponding to the GW. The Hojman constant of motion in terms of Killing vectors also reduces to the GR constant of motion for spinless particles. Substitution of expression (\ref{9}) into the expression (\ref{6}) yields
\begin{equation} 
\frac{du^{\mu}}{ds}+[\frac{1}{2}{\eta}^{{\mu}{\gamma}}(h_{{\gamma}{\alpha},{\beta}}+h_{{\gamma}{\beta},{\alpha}}-h_{{\alpha}{\beta},{\gamma}})+
{K^{\mu}}_{{\alpha}{\beta}}]u^{\alpha}u^{\beta}=0
\label{11}
\end{equation}
This equation shall be used in the next section to investigate the motion of spinless test particles distributed along a ring \cite{13,14} in the field of an incident GW in torsionic background.
\section{The motion of a ring of spinless particles on a $T_{4}$ GW background}
Let us now consider that the background contortion is constant and that the GW hits the ring of spinless particles in $T_{4}$. For a spinless particle initially at rest ,or moving at low speed
$u^{0}=1$ and $u^{k}=0$ the equation (\ref{11}) reduces to
\begin{equation}
\frac{du_{\mu}}{ds}=-[(h_{{\mu}0,0}-\frac{1}{2}h_{00,{\mu}})+K_{{\mu}00}]
\label{12}
\end{equation}
But since for the physical GW we have $h_{{\mu}0,0}=0$ the equation (\ref{12}) reduces \begin{equation}
\frac{du_{\mu}}{ds}=-K_{{\mu}00}
\label{13}
\end{equation}
Here is a very interesting physical departure from GR since there for two particles placed at distance ${\Delta}x=2x_{0}$ is constant which satisfies equation (\ref{13}) in the absence of contortion. Thus in the presence of contortion and taking into account that the distance $l$ between the particles depend also on the metric we obtain
\begin{equation}
{\Delta}l^{2}=-g_{11}{{\Delta}x}^{2}=[1-h_{11}]{\Delta}x
\label{14}
\end{equation}
and
\begin{equation}
{\Delta}x=-K_{x00}t^{2}+At+B
\label{15}
\end{equation}
where A and B are integration constants and $K_{x00}=K$ is the constant contortion. Note that this expression already indicates that the motion of spinless particles is damped by the Cartan contortion effect. The separation of particles in $T_{4}$ spacetime is 
\begin{equation}
{\Delta}l^{2}= [1- R cos{\omega}t](A-Kt^{2}+Bt)
\label{16}
\end{equation} 
where R is the wave amplitude and ${\omega}$ is the wave frequency. Since teleparallelism is equivalent theory to GR we may use the plane wave solution of GR in the expression (\ref{16}) given by $h_{11}= R cos{\omega}t$. Let us now compute the tidal force necessary to produce this motion is
\begin{equation}
\frac{d^{2}{\Delta}l^{2}}{ds^{2}}=-2K{\theta}(t)+\frac{{\omega}^{2}Rcos{\omega}t}{2}[B+At-Kt^{2}]+{\gamma}(t)[A-2Kt]-\frac{R{\omega}}{2}sin{\omega}t[A-2Kt]
\label{17}
\end{equation}
where ${\theta}(t)=[1-Rcos{\omega}t]$ and ${\gamma}(t)=[1+\frac{1}{2}{\omega}Rsin{\omega}t]$. This expression shows that contortion tidal forces appears representing damping forces which contributes for damping the motion of the spinless particles when the ring of spinless particles is hit by a plane GW. It is important to notice that at the $t=0$ the tidal force (\ref{17}) reduces to 
\begin{equation}
\frac{d^{2}{\Delta}l^{2}}{ds^{2}}=(A-2K)(\frac{R{\omega}}{2})
\label{18}
\end{equation}
when we take the Nitsch value for the contortion at the surface of the Earth $10^{-24}s^{-1}$ one obtains a extremely weak torsion contribution to the tidal force. One one the ways to improve this result would be to increase the frequency and amplitude of the GW which from expression (\ref{18}) interacts with the contortion $K$.
\section{Contortion waves and the motion of spinless particles}
In this section we show that the simple definition of teleparallelism as the vanishing of RC curvature may lead us to estimate the contortion in according to same data of GW detectors. We consider here both modes $h_{X}$ and $h_{+}$ of the plane GW.  Since the metric $g_{ij}$ changes in time with the passage of the wave this causes the proper distance between the spinless test particles \cite{13} change as well with the passage with torsion wave. For a plane wave propagating along the $z-axis$, the proper distance in the $xy$-plane \cite{14} is given by
\begin{equation}
dl = [(1+{h^{TT}}_{xx})dx^{2}+ (1-{h^{TT}}_{xx})dy^{2}+2 {h^{TT}}_{xy}dx dy]^{\frac{1}{2}}
\label{19}
\end{equation}
where the metric perturbation components are
\begin{equation}
{h^{TT}}_{xx}= -{h^{TT}}_{yy}=A_{+} e^{-i{\omega}(t-z)}
\label{20}
\end{equation}
\begin{equation}
{h^{TT}}_{xy}= A_{X} e^{-i{\omega}(t-z)}
\label{21}
\end{equation}
The Riemann curvature components are obtained by making use of the $T_{4}$ condition on vanishing of the RC curvature tensor. This yields 
\begin{equation}
{R^{0}}_{1010}=[{\partial}_{1}K_{010} - {\partial}_{0}K_{110}]
\label{22}
\end{equation}
\begin{equation}
{R^{0}}_{1310}=[{\partial}_{3}K_{110} - {\partial}_{1}K_{310}]= -\frac{1}{2}\ddot{h_{+}}= -\frac{{\omega}^{2}}{2}A_{+}e^{-i{\omega}(t-z)}
\label{23}
\end{equation}
\begin{equation}
{R^{0}}_{1313}=[{\partial}_{3}K_{311} - {\partial}_{1}K_{313}] 
\label{24}
\end{equation}
\begin{equation}
{R^{0}}_{2020}=-[{\partial}_{0}K_{220} - {\partial}_{2}K_{020}]
\label{25}
\end{equation}
\begin{equation}
{R^{0}}_{2320}=[{\partial}_{3}K_{220} - {\partial}_{2}K_{320}]
\label{26}
\end{equation}
\begin{equation}
{R^{0}}_{2323}=[{\partial}_{3}K_{223} - {\partial}_{2}K_{323}]
\label{27}
\end{equation}
\begin{equation}
{R^{0}}_{1020}=[{\partial}_{0}K_{120} - {\partial}_{1}K_{020}]= -\frac{1}{2}{\ddot{h}}_{X}
\label{28}
\end{equation}
\begin{equation}
{R^{0}}_{1320}=[{\partial}_{3}K_{120} - {\partial}_{1}K_{320}] 
\label{29}
\end{equation}
\begin{equation}
{R^{0}}_{1023}=[{\partial}_{3}K_{120} - {\partial}_{1}K_{023}]= -\frac{1}{2}{\ddot{h}}_{X}
\label{30}
\end{equation}
\begin{equation}
{R^{0}}_{1323}=[{\partial}_{3}K_{123} - {\partial}_{1}K_{323}]= -\frac{1}{2}{\ddot{h}}_{X}
\label{31}
\end{equation}
Here ${R^{0}}_{{\alpha}{\beta}{\mu}{\nu}}$ represents the Riemann tensor components. Analogous expressions for the $h_{+}$ mode allows us to perform the following computations. Since the computations are analogous we do not make all of them here but is enough to show that it works in some cases. Thus let us consider the case of the component 
\begin{equation}
{R^{0}}_{2020}=-[{\partial}_{0}K_{220} - {\partial}_{2}K_{020}]
\label{32}
\end{equation}
Since by definition $h(t-z)$ for both modes and $(x^{0}=t,x^{1}=x,x^{2}=y,x^{3}=z)$ this expression reduces to
\begin{equation}
{R^{0}}_{2020}=-{\partial}_{0}K_{220}= -\frac{1}{2}\ddot{h_{+}}
\label{33}
\end{equation}
By analogy considering the component 
\begin{equation}
{R^{0}}_{2320}={\partial}_{3}K_{220}= -\frac{1}{2}\ddot{h_{+}}
\label{34}
\end{equation}
From expressions (\ref{32}) and (\ref{33}) we obtain 
\begin{equation}
({\partial}_{0}+{\partial}_{3})K_{220}= 0
\label{35}
\end{equation}
By operating with expression ${\partial}_{0}-{\partial}_{3}$ on the LHS of equation (\ref{34}) one obtains the expression for the wave equation for the contortion component $K_{220}$  
\begin{equation}
({{\partial}^{2}}_{0}-{{\partial}^{2}}_{3})K_{220}= 0
\label{36}
\end{equation}
which can be written as ${\Box}{K_{220}}=0$ , where ${\Box}={\partial}_{\alpha}{\partial}^{\alpha}$ is the D'Alembertian wave operator. This wave equation can be shown to be valid for all other contortion components to yield the following contortion wave equation 
\begin{equation}
{\Box}K_{{\alpha}{\mu}{\nu}}= 0
\label{37}
\end{equation}
Taking for example the component $K_{123}$ one obtains the following complex solution
\begin{equation}
K_{123}= -\frac{i{\omega}}{2}A_{X}e^{-i{\omega}(t-z)}
\label{38}
\end{equation}
Note that this expression shows us that the contortion wave has a phase difference of $\frac{\pi}{2}$ w.r.t the GW. When one takes the real part of the complex phase this expression reduces to
\begin{equation}
Im(K_{123})=- \frac{A_{X}}{2}{\omega}cos({\omega}(t-z))
\label{39}
\end{equation}
where the Im denotes the imaginary part of the complex representation of the torsion wave. When the torsion wave passes by a ring of spinning particles the perturbation of the ring in the $xy$-plane is
\begin{equation}
{\delta}l^{X} = -\frac{A_{X}}{2}l cos({\omega}(t-z))
\label{40}
\end{equation}
Thus contortion contributes to ${\delta}l$ which is the change of the separation of the spinning particles when the torsion wave hits the ring. Actually is better to consider the relative displacement of particles in the ring of particles which is given by $\frac{{\delta}l}{l}$ which by the expressions (\ref{39}) and (\ref{40}) yields
\begin{equation}
|Im(K_{123})|= {\omega}\frac{{\delta}l^{X}}{l}
\label{41}
\end{equation}
Note that by considering a frequency of the order of ${\omega}=10^{-3}Hz$ and the value obtained by Nitsch \cite{12} for the contortion at the surface of the Earth of $K_{123}=10^{-24} s^{-1}$ one obtains from expression (\ref{41}) a relative displacement for the ring of 
\begin{equation}
\frac{{\delta}l^{X}}{l}=10^{-21}
\label{42}
\end{equation}
which is of the order of the gravitational wave Peres gedanken experiment result. Let us now consider $T_{4}$ amplitude of perturbations of spacetime in terms of contortion in general case to express the analogous of the gravitational field energy momentum tensor. This model yields the following Riemann curvature tensor w.r.t contortion
\begin{equation}
R^{a}_{{0}{b}{0}}(g)={\partial}^{a}{K_{0b 0}}-{\partial}_{0}{{K}^{a}}_{b 0}
\label{43}
\end{equation}
where here g is the symbolic representation of the metric tensor $g_{{\mu}{\nu}}$. By considering that the contortion is just a function of time and that the metric curvature can be expressed as 
\begin{equation}
R_{a0b0}(g)=-\frac{1}{2}{{\ddot{h}}^{TT}}_{ab}= -{\partial}_{0}{K_{ab0}}
\label{44}
\end{equation}
which allows us to write the contortion in terms of the acceleration of metric perturbation h as
\begin{equation}
\frac{1}{2}{{\dot{h}}^{TT}}_{ab}= {K_{ab0}}
\label{45}
\end{equation}
which allows us to find the tensor or GW perturbation in terms of the Cartan contortion tensor as
\begin{equation}
{{h}^{TT}}_{ab}= 2\int{{K_{ab0}}dt}
\label{46}
\end{equation}
Substitution of this last expression into the expression for the pseudo-tensor for the GW field 
\begin{equation}
{t^{GW}}_{{\mu}{\nu}}=\frac{1}{32{\pi}}<{{h}^{TT}}_{ab,{\mu}}{{{h}^{TT}}^{ab}}_{,{\nu}}>
\label{47}
\end{equation}
yields
\begin{equation}
{t^{GW}}_{{0}{0}}=\frac{1}{32{\pi}}<{{h}^{TT}}_{ab,0}{{h}^{TT}}^{ab}_{,{0}}>
\label{48}
\end{equation}
or
\begin{equation}
{t^{GW}}_{00}=\frac{1}{8{\pi}}< K^{2}>
\label{49}
\end{equation}
where $K^{2}= K_{ab0}K^{ab0}$ where $a,b=1,2,3$. This expression could be also obtained directly from the expression derived by de Andrade et al. \cite{4} for the energy-density of the gravitational field
\begin{equation}
{t}_{{\mu}{nu}}=h^{A}_{\mu} j_{A{\nu}}+\frac{c^{4}}{4{\pi}}{{\Gamma}^{\alpha}}_{{\mu}{\beta}} {{S_{\alpha}}^{\beta}}_{\nu}
\label{50}
\end{equation}
in the absence of gauge current $j^{\mu}_{A}$ where tetrad indices are represented by latin capital letters. Actually is is not difficult to show that contortion contribution to the gravitational field energy (\ref{48}) is contained into expression (\ref{50}). This can be obtained as follows. Consider that the $00-component$ of (\ref{50}) can be expressed as
\begin{equation}
{t}_{{0}{0}}=h^{A}_{0} j_{A{0}}+\frac{c^{4}}{4{\pi}}{{\Gamma}^{\mu}}_{{0}{\nu}}{{S_{\mu}}^{\nu}}_{0}
\label{51}
\end{equation}
and since 
\begin{equation}
S^{{\mu}{\nu}0} = \frac{1}{2}[K^{{\mu}{\nu}0}-g^{{\mu}0}{{T^{\alpha}}_{\alpha}}^{\nu} +g^{{\mu}0}{T^{{\beta}{\nu}}}_{\beta}] 
\label{52}
\end{equation}
where $T^{{\mu}{\nu}{\alpha}}$ is the torsion tensor we notice that the second term of expression (\ref{50}) reduces to the contortion squared expression. A more detailed discussion of the possibility of designing these kind of experiments as well as a detector of spin polarized to test torsion theories of gravity where torsion propagates were previously discussed by Hojmann, Rosembaum and Ryan (HRR) \cite{15}. Actually HRR have suggested that it is possible to find an energy flow required to excite a cold detector in nearly steady-state vibrations possesing energies at room temperature thermal vibrations. Taking into account a detector of Weber type , for a frequency of the order $10^{3} Hz$ torsion would have to be of the order of $10^{21} cm^{-1}$ which is an unrealistic value that has never been observed. Thus HRR have reached the conclusion that GW experiments that have not detected excitations of this magnitude, actually gives us the information that this huge number could be taken as a rough upper limit for torsion. Since this number seems to be extremely high even for very massive astrophysical sources such as black holes and binary pulsars, our model seems to be much more realistic which again shows that teleparallel gravity even today maybe useful not only to explain important physical problems such as the issue of the gravitational field energy but also help us to understand some pratical problems such as torsion and gravitational waves. Other interesting application is the generalization of Nieto and Ryan work \cite{8} of the motion of spinning particles on a plane GW to $T_{4}$. Other interesting terrestrial proposals to torsion detection have been recently appeared in the literature by Lammerzahl \cite{16} and myself \cite{17}.

\newpage

\section*{Acknowledgements}
 I would like to express my gratitude to Prof. I.D. Soares, Prof. M. Novello and Prof. J.G.Pereira for pointing out an error in the first draft of this paper. Thanks are also due to Dr. Vanessa de Andrade for helpful discussions on the subject of this paper. Partial financial support from CNPq. is grateful acknowledged.


\begin{thebibliography}{17}
\bibitem{1} A. Einstein,Mathematische Annalen (1930) 102 Band 5,685. 
\bibitem{2} W. Adamowicz, General Relativity and Gravitation J,vol.12,No.9(1980) 677.
\bibitem{3} R. Sippel and H. Goenner, General Relativity and Gravitation J,Vol.18,No.12,(1986)1229.
\bibitem{4} V. de Andrade,L.C. Guillen, J.G. Pereira, Phys. Rev. Lett. (2001).
\bibitem{5} S. Hojman,Phys. Rev. D(1978).
\bibitem{6} F. W. Hehl,Phys. Lett. 36 A (1971),225.
\bibitem{7} H. Kleinert, Gauge Fields and Condensed Matter: Differential Geometry (1989) World Scientific.
\bibitem{8} J.A. Nieto and M. P. Ryan jr.,Il Nuovo Cimento 63 A,1 (1981) 71. 
\bibitem{9} I. L. Shapiro, Physical Aspects of space-time torsion (2002) 357,2,115.
\bibitem{10} R. Hammond, Torsion Gravity (2002) Reports on Progress in Physics.
\bibitem{11} L.L. Smalley, Phys. Rev. D (1978).
\bibitem{12} J. Nitsch in Cosmology and Gravitation:Spin,Torsion,Rotation and Supergravity (1980).Eds. P.G.Bergmann and V. de Sabbata,Plenum Press.
\bibitem{13} I.Ciufolini and J.A.Wheeler,Inertia and Gravitation,(1995) Princeton University Press.
\bibitem{14} H. Ohanian, Gravitation and Spacetime.
\bibitem{15} S. Hojman, M. Rosenbaum and M. P. Ryan, Phys. Rev. D 19,2 (1978) 430. 
\bibitem{16} C.L\"{a}mmerzahl,Phys.Lett.228A (1997)223.
\bibitem{17} L.C. Garcia de Andrade, Class. and Quantum Gravity (2001) 18,18,3907.
\end{thebibliography}
\end{document}